\documentclass[12pt,fleqn]{article} 
\usepackage{graphics} 
\usepackage{cite}
\usepackage{amsfonts} 

\newcommand{\gsim}{\stackrel{\scriptstyle >}{{ }_{\sim}}}

\newcommand{\mb}{\ensuremath{m_b}}
\newcommand{\mt}{\ensuremath{m_t}}
\newcommand{\Dmb}[1][]{\ensuremath{\Delta\mb^{#1}}}
\newcommand{\tb}[1][]{\ensuremath{\tan^{#1}\!\beta}}

\newcommand{\mg}{\ensuremath{M_{\tilde{g}}}}
\newcommand{\msb}[1]{\ensuremath{M_{\tilde{b}_{#1}}}}
\newcommand{\mst}[1]{\ensuremath{M_{\tilde{t}_{#1}}}}
\newcommand{\mstau}[1]{\ensuremath{M_{\tilde{\tau}_{#1}}}}
\newcommand{\msf}[1]{\ensuremath{M_{\tilde{f}_{#1}}}}

\newcommand{\TeV}{\ensuremath{\,{\rm TeV}}}
\newcommand{\GeV}{\ensuremath{\,{\rm GeV}}}

\newcommand{\rsm}{\ensuremath{R^{\rm SM}}}
\newcommand{\rmssm}{\ensuremath{R^{\rm MSSM}}}
\newcommand{\brHbb}{\ensuremath{BR(H\to b\bar{b})}}
\newcommand{\brHtt}{\ensuremath{BR(H\to \tau^+\tau^-)}}
\newcommand{\brhott}{\ensuremath{BR(h \to \tau^+\tau^-)}}
\newcommand{\mtau}{\ensuremath{m_\tau}}
\newcommand{\mbQ}{\ensuremath{\mb(Q)}}
\newcommand{\mbsH}{\ensuremath{\mb^2(M_H)}}
\newcommand{\hb}{\ensuremath{h_b}}

\newcommand{\sa}{\ensuremath{\sin\alpha}}
\newcommand{\ca}{\ensuremath{\cos\alpha}}
\newcommand{\ta}{\ensuremath{\tan\alpha}}
\newcommand{\sbt}{\ensuremath{\sin\beta}}
\newcommand{\cbt}{\ensuremath{\cos\beta}}
\newcommand{\ma}{\ensuremath{M_{A^0}}}
\newcommand{\mz}{\ensuremath{M_{Z}}}

\newcommand{\Dmtau}{\ensuremath{\Delta m_\tau}}
\begin{document} 
\begin{titlepage} 
\hfill{} 
\begin{tabular}{l} 
KA-TP-19-2001 \\ 
hep-ph/0106027 \\ 
June 2001 \\
\vspace*{0.4cm}
\end{tabular} 
\begin{center}
\textbf{\large Distinguishing Higgs models 
in $H\to b\bar{b} / H\to \tau^+\tau^-$}
\vspace*{0.7cm}\\
\renewcommand{\thefootnote}{\fnsymbol{footnote}}
{\par\centering 
Jaume Guasch~\footnote{Supported by 
the European Union under contract Nr. HPMF-CT-1999-00150.}, Wolfgang Hollik, 
Siannah Pe{\~n}aranda~\footnote{Supported by 
 \textit{Fundaci{\'o}n Ram{\'o}n Areces}.\\
electronic address: guasch@particle.physik.uni-karlsruhe.de, 
hollik@particle.physik.uni-karlsruhe.de,
siannah@particle.physik.uni-karlsruhe.de}
\par} 
{\par\centering 
\textit{Institut f\"{u}r Theoretische Physik, Universit\"{a}t}
\textit{Karlsruhe,\\  Kaiserstra\ss{}e 12, D--76128 Karlsruhe, Germany }}
\end{center}
\vspace*{0.7cm}
{\par\centering\textbf{\large Abstract}\vspace*{0.5cm}\\ 
\par} 
\noindent We analyze the ratio of branching ratios $R=\brHbb/\brHtt$ 
of Higgs boson decays as
a discriminant quantity between supersymmetric  and
non-supersymmetric models. This ratio receives large
renormalization-scheme independent radiative corrections in
supersymmetric models at large $\tb$,
which are absent in the 
Standard Model or Two-Higgs-doublet models. These corrections are
insensitive to the supersymmetric mass scale.
A detailed analysis in the
effective Lagrangian approach shows that, with a measurement of
$\pm21\%$ accuracy, the Large Hadron Collider can discriminate between
models if the CP-odd Higgs boson mass is below  $900\GeV$. An $e^+e^-$
Linear Collider at $500\GeV$ center of mass energy can discriminate
supersymmetric models up to a CP-odd Higgs mass of $\sim 1.8\TeV$.

\vspace*{1cm}
\end{titlepage} 
\renewcommand{\thefootnote}{\arabic{footnote}}
\setcounter{footnote}{0}

The existence of the scalar Higgs boson of the 
 Standard Model (SM) is still
waiting for experimental confirmation. Last LEP results, suggesting
a light neutral Higgs particle with a mass 
about $115\GeV$ are encouraging~\cite{LEP115},
but we will have to wait the news from the hadron colliders, 
the upgraded Fermilab Tevatron or
the upcoming Large Hadron Collider (LHC) at CERN, to see this result
either confirmed or dismissed. For intermediate masses above the LEP
limit and below $180\GeV$ there is a chance for the
Tevatron~\cite{HiggsRunII}, but for higher masses up to $1\TeV$ one
needs the LHC\cite{ATLASCMS}. However, even if a neutral scalar boson is
discovered, the question 
will still be open: whether it is the Higgs particle of the
minimal Standard Model (SM) 
or whether there is an extended Higgs structure beyond the SM.
In this
paper we approach this question by investigating the neutral Higgs sector
of various types of models. 
In many extensions of the SM the Higgs sector is enlarged,
containing several neutral Higgs bosons as well as charged
ones~\cite{Hunter}. 
At present, supersymmetric (SUSY) models
have become the theoretically favored scenarios, with the
Minimal Supersymmetric Standard Model (MSSM) as the most-predictive
framework beyond the SM~\cite{MSSMreps}.  
The Higgs
sector of the MSSM contains two Higgs doublets. Its properties at the
tree-level are determined by just two free parameters, 
conventionally chosen as the ratio of the
vacuum expectation values (\textit{vev}s) of each doublet, 
$\tb=v_2/v_1$, and the mass of 
the CP-odd neutral Higgs boson, \ma. 
This simple structure is known to receive
large radiative corrections, which have been computed up to
two-loop order~\cite{CHHHWW}; a definite prediction is the existence of a
light neutral scalar boson with mass below $130\GeV$. 
It is also well known that the SUSY one-loop corrections 
to the tree-level couplings of Higgs bosons to bottom quarks can be 
significant for large values of $\tan \beta$,
and that they do not decouple in
the limit of a heavy supersymmetric 
spectrum~\cite{CGGJS,CJS,eff,CMW,Kolda,Pierce,Haberetal,MJHtb},
opposite to their behaviour in electroweak gauge boson 
physics~\cite{Dobadoetal}.

These
one-loop corrections can be translated directly into a redefinition of
the relation between the $b$-quark Yukawa coupling 
(entering production and decay processes)
and the physical (pole) mass of the $b$-quark, with
important phenomenological implications e.g.\ for the branching ratios 
of SUSY Higgs-boson decays into heavy fermions.

Following this path we consider in this letter the ratio of
branching ratios of a neutral Higgs boson $H$,
\begin{equation}
  \label{eq:Rdef}
  R=\frac{\brHbb}{\brHtt}\,\,,
\end{equation}
analyzing in detail the Yukawa-coupling effects
and their phenomenological consequences.
In the SM, after accounting for the leading QCD corrections, one has
$
\rsm=3\,\mbsH/\mtau^2
$, 
where \mbQ\  is the $b$-quark running mass 
based on the QCD evolution,
and $\mtau$ is the $\tau$-lepton mass. 
For  $M_H=115\GeV$ we have $\rsm\simeq8$.
Some other, small, QCD contributions are neglected here.
Actually, the result  for the leading QCD corrections
is much more general in the sense that it is
valid for any Higgs model in
which the Higgs sector follows the family structure of the SM, like the
Two-Higgs-Doublet-Model (2HDM) of type I and II, 
or the MSSM as far as the Standard QCD correction is considered.

The ratio (\ref{eq:Rdef}) is very interesting from both 
the experimental and the theoretical side.
It is a clean observable, measurable in a
counting experiment, with only small 
systematic errors since most of them 
are canceled in the ratio. 
The only
surviving systematic effect results from the efficiency of
$\tau$- and $b$-tagging. 
From the theoretical side,
it is  independent of the production
mechanism of the decaying neutral Higgs boson and of the total 
width; hence,  new-physics effects affecting the production cross-section do
not appear in the ratio~(\ref{eq:Rdef}). For the same reason, 
this observable is insensitive to unknown
high order QCD corrections to Higgs boson production. 

Another theoretical point of view is of interest: 
When one finds large radiative corrections to a certain
process (e.g.\ $H\rightarrow b \bar b$), one may wonder if their 
effects would be 
absorbed by a proper redefinition of the parameters 
in some renormalization scheme,
such that these effects disappear.
Since the ratio~(\ref{eq:Rdef}) only depends on the ratio of the masses,
  there is no other parameter (e.g.\  \tb) that could absorb
these large corrections. 

The partial decay width $\Gamma(h\to b \bar{b})$ 
of the lightest supersymmetric neutral Higgs particle
has been the subject of several studies in the literature. 
Besides the complete
one-loop corrections~\cite{Dabelstein}, 
comprehensive studies of the one- and two-loop SUSY-QCD corrections are
available in Ref.~\cite{CJS} and~\cite{hbbtwoloops}, respectively. 
Implications for 
Higgs-boson searches from SUSY effects in the
$hb\bar{b}$ vertex (together with their effective Lagrangian description)
can be found in~\cite{eff,CMW}.
The decoupling properties
of the SUSY-QCD corrections to $\Gamma(h\to b \bar{b})$ have been
extensively discussed in~\cite{Haberetal}.
The effects on $\brhott$ were presented in~\cite{George}. Analyses 
of the observable $R$ can be found in~\cite{Kolda,Troconiz}.

In the MSSM, the Higgs boson couplings to down-type fermions receive
large quantum corrections, enhanced by \tb. 
In the case of the $t\bar{b}H^+$ vertex, these corrections
have been resummed to all orders of perturbation theory with the help of
the effective Lagrangian formalism in Ref.~\cite{eff}.
The effective Lagrangian of the
MSSM Higgs couplings to down-type fermions can be written as follows,
\begin{equation}
  \label{eq:defEffL}
  {\cal L}_{\rm eff}=\hb\left( -\varepsilon_{ij} H_1^{i} L^{j} B_R +
    \Delta_B H_2^i L^{i} B_R\right) ,
\end{equation}
where $H_1$ and $H_2$ are the two Higgs-doublets of the MSSM, 
$L$ is the $SU(2)_L$ fermion-doublet,
$B_R$ is the right-handed down-type fermion,\footnote{Here, and in
  the following, we use the third generation quark notation as a generic
  one.} and  \hb\  is
the $b$-quark Yukawa coupling, related to the 
corresponding running mass at the
tree level by
$\hb=\mb/v_1$. 
$H_2$ is
the doublet responsible for giving masses to the up-type fermions, and
the second term in~(\ref{eq:defEffL}) 
only appears when radiative corrections are taken into account
(encoded in the quantity $\Delta_B$)
due to breaking of SUSY. 
On the other hand,
in the most general 2HDM such
terms are permitted also at the tree level. 
However, they would lead to large Flavour
Changing Neutral Currents in the light-quark sector of the model, and
hence they are usually explicitly forbidden by a postulated
ad-hoc symmetry,
which leads to the so called 2HDM of Type I and Type II.

Given the effective Lagrangian~(\ref{eq:defEffL}), 
with the \textit{vev} $v_i$ of the Higgs doublet  $H_i$,
the $b$-quark mass is given by~\footnote{Notice that in the case of vanishing
  tree-level Yukawa coupling, the bottom quark mass would be gene\-rated by the
  non-decoupling terms like $\Delta_B$ 
in~(\ref{eq:deffmb})~\cite{Borzumatti}.} 
\begin{equation}
  \label{eq:deffmb}
  \mb=\hb(v_1+\Delta_B\, v_2)=\hb v_1 \left(1+\Delta_B \,\tb\right)\equiv\hb
  v_1 \left(1+\Delta\mb\right)\,. 
\end{equation}
We now can relate the known quark mass to the Yukawa coupling via
\begin{equation}
  \label{eq:deffhb}
  \hb=\frac{\mb}{v_1} \frac{1}{1+\Dmb}=
      \frac{\mb}{v \cos\beta}\frac{1}{1+\Dmb}, \quad \quad
    v= (v_1^2+v_2^2)^{1/2} \, .
\end{equation}
$\Dmb$ is a non-decoupling quantity that encodes the leading radiative
corrections. 
The expression~(\ref{eq:deffmb})
contains the resummation of all possible $\tb$
enhanced corrections of the type 
$(\alpha_{(s)} \tb)^n$~\cite{eff}. 
Similarly to the $b$ case, the relation between $m_\tau$ and 
the $\tau$-lepton Yukawa coupling $h_\tau$ is also modified
by a quantum correction $\Dmtau$, in analogy to~(\ref{eq:deffhb}). 

\begin{figure}
\centerline{\resizebox{10cm}{!}{\includegraphics{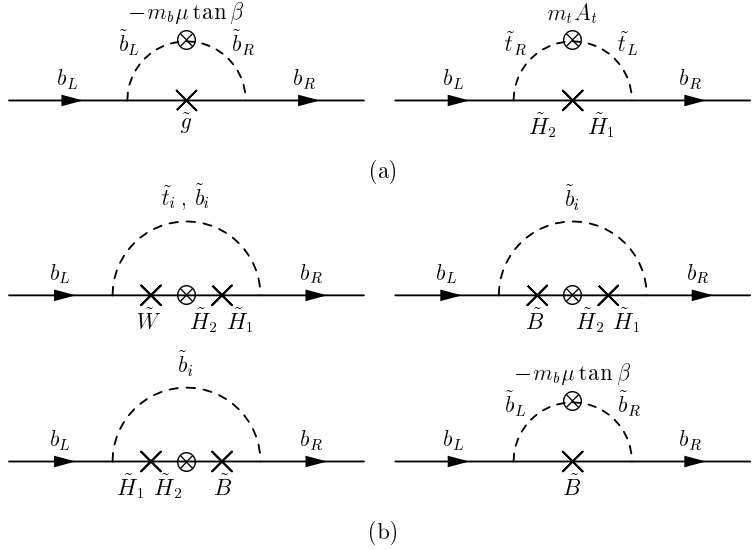}}}
\vspace*{-0.2cm}
\caption{Diagrams contributing to $\Dmb$ -- eq.~(\ref{deltamb}). The
  cross means a mass insertion, and the cross with a circle the coupling
  with $H_2$. The diagrams contributing to $\Dmtau$ are those equivalent
  to \textbf{(b)}.}\label{fig:leadingdiagrams}
\end{figure}
The explicit form of $\Dmb$ and $\Dmtau$ at the one-loop level can be 
obtained approximately 
by computing the supersymmetric loop diagrams at zero external momentum
($M_{SUSY} \gg m_b\,,m_\tau$), as given in
Fig.~\ref{fig:leadingdiagrams} for $\Dmb$. The 
dominant diagrams are those of 
Fig.~\ref{fig:leadingdiagrams}a, but in order
to have a precise evaluation  
we are including the entire set in our result, which is 
given by
\begin{eqnarray}
\Dmb &\simeq &\mu\tb\,\Bigg\{
{2\alpha_S \over 3\pi} M_{\tilde g}\,I(M_{\tilde b_1},
M_{\tilde b_2},M_{\tilde g})
 + {Y_t \over 4\pi} A_t\,I(M_{\tilde t_1},M_{\tilde t_2},\mu)
\nonumber\\
&+&\frac{\alpha}{4\pi}\Bigg(-\frac{ M_2}{s_W^2}\bigg( \left[
c_t^2~I(M_{\tilde t_1},M_2,\mu)+s_t^2~I(M_{\tilde t_2},M_2,\mu)
\right]\nonumber\\
&&
+\frac{1}{2}\left[
c_b^2~I(M_{\tilde b_1},M_2,\mu)+s_b^2~I(M_{\tilde b_2},M_2,\mu)
\right]\bigg)\nonumber\\
&&
-\frac{M_1}{3\,c_W^2}\bigg(
\frac{1}{3}I(M_{\tilde b_1},M_{\tilde b_2},M_1)\nonumber
+\frac{1}{2}\left[
c_b^2~I(M_{\tilde b_1},M_1,\mu)+s_b^2~I(M_{\tilde b_2},M_1,\mu)
\right]\nonumber\\
&&+\left[
s_b^2~I(M_{\tilde b_1},M_1,\mu)+c_b^2~I(M_{\tilde b_2},M_1,\mu)
\right]\bigg)\Bigg)\Bigg\}\,.\label{deltamb}
\end{eqnarray}
The diagrams contributing to $\Dmtau$ are those equivalent to
Fig.~\ref{fig:leadingdiagrams}b replacing $b\to\tau$,
$\tilde{b}\to\tilde{\tau}$, $\tilde{t}\to\tilde{\nu}_\tau$. 
Explicitly, they read
\begin{eqnarray}
\Dmtau &\simeq & \mu\tb\frac{\alpha}{4\pi}\Bigg\{
-\frac{M_2}{ s_W^2}\left(
I(M_{\tilde\nu_\tau},M_2,\mu)
+\frac{1}{2} \left[
c_{\tau}^2~I(M_{\tilde \tau_1},M_2,\mu)+
s_{\tau}^2~I(M_{\tilde \tau_2},M_2,\mu)
\right]\right)\nonumber\\
&&+\frac{M_1}{c_W^2}\bigg(
I(M_{\tilde\tau_1},M_{\tilde\tau_2},M_1)
+\frac{1}{2}\left[
c_\tau^2~I(M_{\tilde \tau_1},M_1,\mu)+s_\tau^2~I(M_{\tilde \tau_2},M_1,\mu)
\right]\nonumber\\
&&-\left[
s_\tau^2~I(M_{\tilde \tau_1},M_1,\mu)+c_\tau^2~I(M_{\tilde \tau_2},M_1,\mu)
\right]\bigg)\Bigg\}\,.\label{deltamtau}
\end{eqnarray}
In the above expressions we have introduced shorthand notations 
for the functions
of the Weinberg angle $s_W \equiv \sin\theta_W$, $c_W \equiv
\cos\theta_W$, the top quark Yukawa coupling
$Y_t = \frac{g^2 m_t^2}{8\pi m_W^2 \sin\beta^2}$, and  sine and
cosine of the sfermion mixing angles $s_{t,b}\,, c_{t,b}\,,$ 
and $s_\tau, c_\tau$. 
For further conventions and notation see Refs.~\cite{CGGJS,Dobadoetal}.
The fine structure constants, $\alpha_S$ and $\alpha$, have to be
evaluated at the SUSY mass scale.
The function 
$I$ is given by,
\begin{eqnarray}
I(a,b,c) = {a^2b^2\ln(a^2/b^2)+b^2c^2\ln(b^2/c^2)+c^2a^2\ln(c^2/a^2) \over
(a^2-b^2)(b^2-c^2)(a^2-c^2)}.
\end{eqnarray}
Although partial results for the expressions~(\ref{deltamb}),
  (\ref{deltamtau}) have been given several times in the
  literature~\cite{eff,CMW,Kolda,Pierce}, 
the subleading terms have 
not been given so far in a complete 
version.{\footnote{Notice a sign difference in $\Dmtau$ 
with~\cite{CMW}.}} We have checked  the results 
in~(\ref{deltamb}), (\ref{deltamtau})
using {\it FeynArts 3} and
{\it FormCalc}~\cite{Hahn}.

If we impose that all the SUSY masses, and also the supersymmetric Higgs mass
parameter $\mu$,
are approximately of the same scale, $M_{SUSY}$,
\[\hspace*{1.3cm}
M_{\tilde f}\, (\tilde f \equiv \tilde t,\tilde b,\tilde \tau,\tilde \nu) 
\sim M_g \sim M_1 \sim M_2 \sim \mu \sim M_{SUSY}\,,
\]
we find that:
\begin{eqnarray}
  \Dmb&\simeq& {\rm sign}(\mu)\tb\left\{
\frac{\alpha_S }{3 \pi}- \frac{\alpha}{16 \pi s_W^2} 
\left(3+\frac{11}{9} \frac{s_W^2}{c_W^2}\right)
+\frac{Y_t}{8\pi}\,\frac{A_t}{M_{SUSY}}\,\right\}\,,\nonumber\\
\Dmtau&\simeq&  -\, {\rm sign}(\mu)\tb\, \frac{\alpha}{16 \pi s_W^2}\,
\left(3-\frac{s_W^2}{c_W^2}\right)\,\,. 
\label{eq:dmblimit}
\end{eqnarray}
Notice that these two quantities are independent of the SUSY
mass scale $M_{SUSY}$ since they only depend on $\tb$ and 
the ratio $A_t/M_{SUSY}$.

From the effective Lagrangian~(\ref{eq:defEffL}), the
$b$-quark coupling to each of the MSSM neutral Higgs bosons~\cite{CMW}
is also derived:
\begin{eqnarray}
  h^0 b\bar{b}&:& C_{hbb}=\hb \sa
  \left(1-\frac{\Dmb}{\tb\ta}\right)=\frac{\mb\sa}{v \cbt} \Delta_{hbb}\ \ , \ \ \nonumber\\
  H^0 b\bar{b}&:& C_{Hbb}=-\hb \ca
  \left(1+\frac{\Dmb\ta}{\tb}\right)=-\frac{\mb \ca}{v \cbt} \Delta_{Hbb}\ \ , \ \ \nonumber\\
  A^0 b\bar{b}&:& C_{Abb}=-i \hb \sbt \left(1-\frac{\Dmb}{\tb^2}\right)=-i
  \mb \tb \Delta_{Abb}  \label{eq:DeffCoupH}\,\,.
\end{eqnarray}
Notice that, although $\Dmb$ is basically a non-decoupling quantity, 
the CP-even mixing angle behaves as $\ta\to-1/ \tb$ in the
\textit{decoupling regime} of the MSSM Higgs sector (i.e.\ $\ma\gg\mz$)
and the $h^0b\bar{b}$
coupling reaches the SM value $C_{hbb}\to \mb/v$. 
A very detailed analysis of this decoupling behaviour (at
  one-loop order) can be found in Ref.~\cite{Haberetal}.

Now we analyze the deviation of the ratio~(\ref{eq:Rdef}) 
from the SM value, 
caused by the SUSY radiative corrections,
for each  
of the MSSM neutral Higgs bosons $\phi=h,H,A$, in terms of the quantity
\begin{equation}
  \label{eq:AnaliR}
  \frac{\rmssm(\phi)}{\rsm}=\frac{1}{\rsm}\frac{3 C_{\phi
  bb}^2}{C_{\phi\tau\tau}^2}=\left(\frac{\Delta_{\phi bb}}{\Delta_{\phi\tau\tau}}\right)^2\,\,,
\end{equation}
which is a function depending only on $\tb$, $\ta$, $\Dmb$ and
$\Dmtau$,  and encoding all the genuine SUSY corrections.
The contributions from QCD 
are the same as in the SM, and they cancel
in~(\ref{eq:AnaliR}). 
Differences in the electroweak corrections can occur only from  
loops with Higgs particles, and  they can usually be 
neglected. In a MSSM-like Higgs sector, the Higgs-boson
loop contributions are very small compared to the rest of the
corrections. Large corrections from the Higgs boson
sector can only 
arise in models in which the splitting between the 
Higgs bosons masses is much larger than that
of the MSSM.\footnote{A similar situation in the  $t\bar{b}H^-$
    coupling can be seen comparing the SUSY Higgs sector
    contributions~\cite{CGGJS} with the  
    2HDM ones~\cite{CGHS}.} If this situation were to be found in the
experiments, the 
MSSM would be excluded without any further analysis. Moreover, the
contributions from the Higgs sector are very similar for the $\phi bb$
and $\phi\tau\tau$ vertices, and they will mostly cancel in
$\rmssm(\phi)$. The genuine SUSY corrections, on
the other hand, present sizeable differences 
between the $\phi bb$ and $\phi\tau\tau$
couplings, even in the case of similar squark and slepton spectra:
\begin{itemize}
\item the SUSY-QCD corrections mediated by gluinos 
is only present in $\Dmb$ [1st term in~(\ref{deltamb})], 
yielding the by
far dominant contribution to~(\ref{eq:AnaliR});
\item there exists a contribution from the
chargino sector to $\Dmb$ resulting from mixing in the stop sector [2nd term
in~(\ref{deltamb})], whereas a corresponding term is not present 
in $\Dmtau$  due to the absence sneutrino mixing;
\item the contribution from the $\tilde{B}$ loops is
different in both cases because of the different hypercharges.
\end{itemize}

In the following we concentrate on the case of
the lightest CP-even Higgs
boson, $h^{0}$. 
The ratio $R$ defined in~(\ref{eq:Rdef}), written in terms of
the non-decoupling quantities $\Dmb$ and $\Dmtau$ and normalized to the
SM value, reads
\begin{equation}
\label{eq:Rh0}
\frac{\rmssm(h)}{\rsm}=\frac{(1 + \Dmtau)^2\,(-\cot\alpha \Dmb + \tan \beta)^2}
{(1 + \Dmb)^2\,(-\cot\alpha \Dmtau + \tan \beta)^2}\,.
\end{equation}
\begin{figure}[t]
\begin{center}
\begin{tabular}{cc}
\resizebox{6cm}{!}{\includegraphics{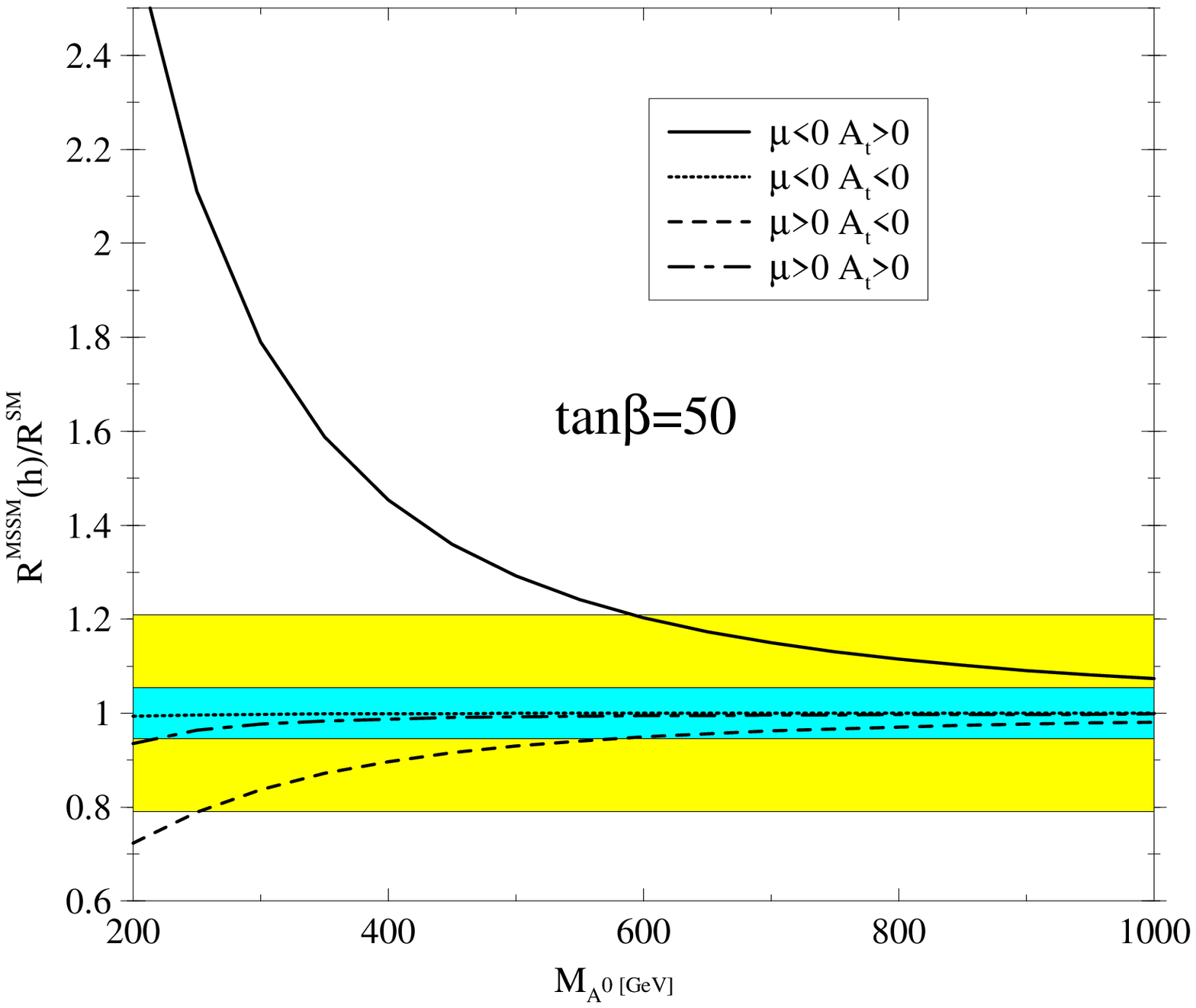}}&
\resizebox{6cm}{!}{\includegraphics{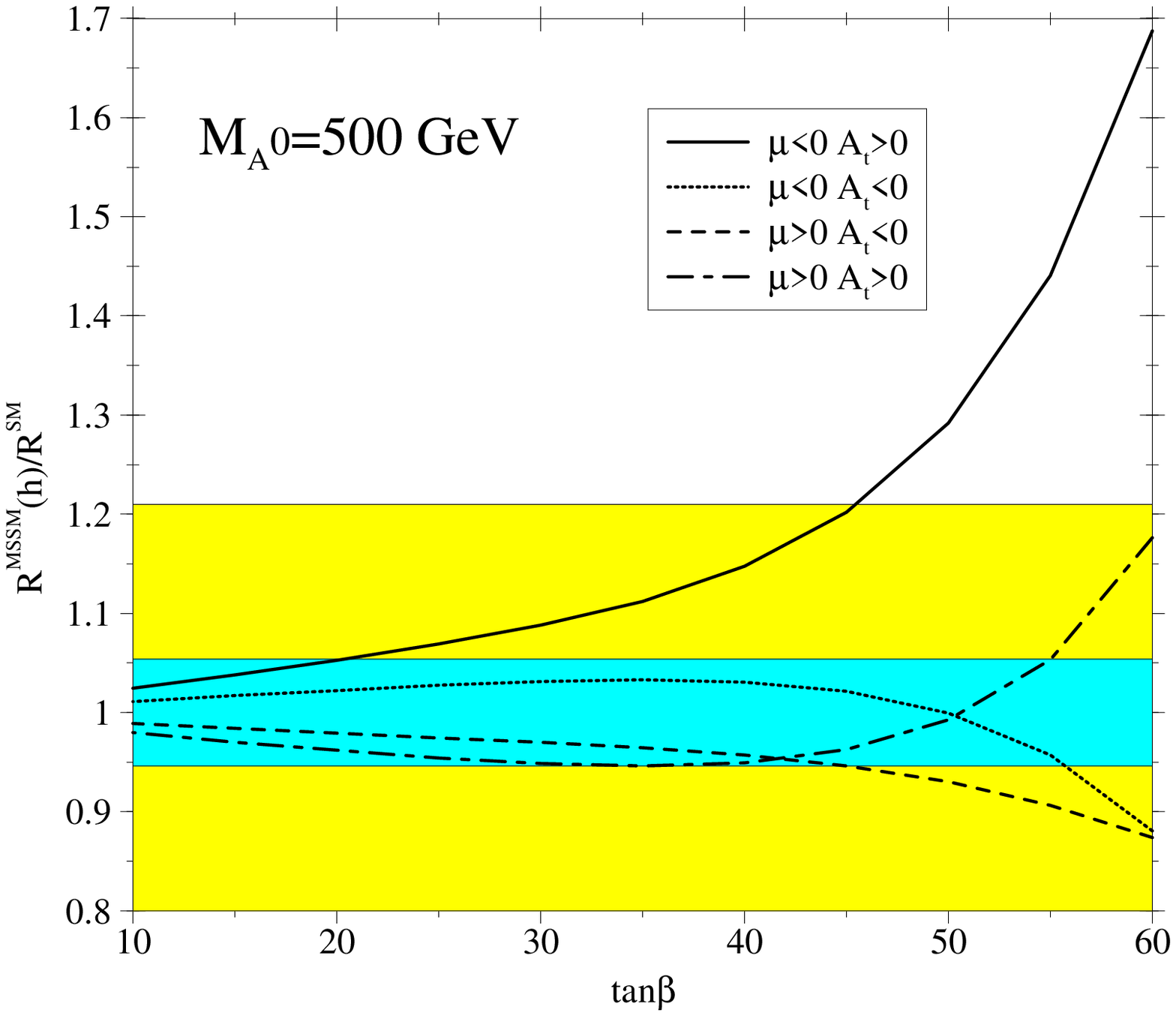}}\\
(a)&(b)
\end{tabular}
\end{center}\vspace*{-0.4cm}
\caption{Deviation of $\rmssm(h)$ with respect to the SM value, as a
  function of \textbf{a)} \ma, and \textbf{b)} \tb, for various choices
  of the SUSY parameters. The light-shaded region shows the $\pm21\%$
  deviation with respect to the SM, and the dark-shaded one the
  $\pm5.4\%$.}
\label{fig:rmssmh}
\end{figure}
In Fig.~\ref{fig:rmssmh} we present numerical results for the
expression~(\ref{eq:Rh0}). The SUSY spectrum has been taken to be around
$1.5\TeV$, namely, 
\[
\mg=\msb1=\mst1=\mstau1=M_2=|\mu|=A_b=A_\tau=|A_t|=1.5\TeV\,\,,
\]
and we assume the usual GUT relation $M_1=5/3 M_2 s^2_W/c^2_W$ and maximal
mixing in the $\tilde{b}$  and $\tilde{\tau}$ sector, $\theta=\pm\pi/4$. 
Our convention here is $\msf1 <\msf2$.
The rest of
the parameters are fixed by the $SU(2)_L$ symmetry. As a consequence, a
certain splitting of order $\sim15\%$ is generated in the sfermion
sector. Nevertheless the approximate expressions~(\ref{eq:dmblimit})
give an accuracy better that $10\%$ in $\Dmtau$ and in $\Dmb$ for
$A_t>0$. For $A_t<0$ the approximation for $\Dmb$ 
is much worse, giving deviations
of $\sim 23\%$ for large $\tb$. 
For definiteness, we also list the following values used for the SM
parameters: $\mt=175\GeV$, $\mb=4.62\GeV$, $\mtau=1.777\GeV$~\cite{PDG}.  
The CP-even
mixing angle is computed including the leading 
corrections up to two-loop order by means of the  
program \textit{FeynHiggsFast}~\cite{FeynHiggsFast}.
\begin{figure}[t]
\begin{center}
\begin{tabular}{cc}
\resizebox{6cm}{!}{\includegraphics{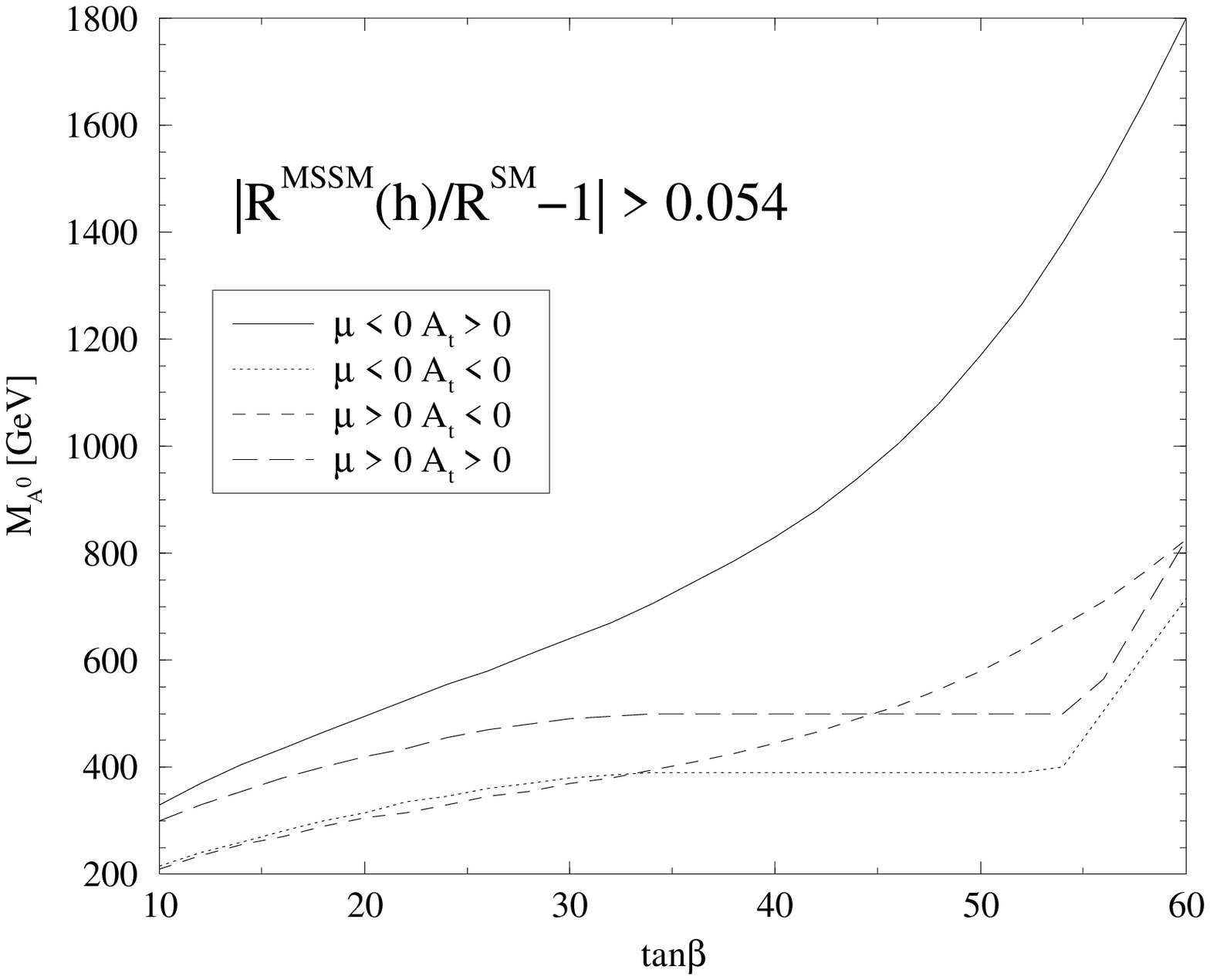}}&
\resizebox{6cm}{!}{\includegraphics{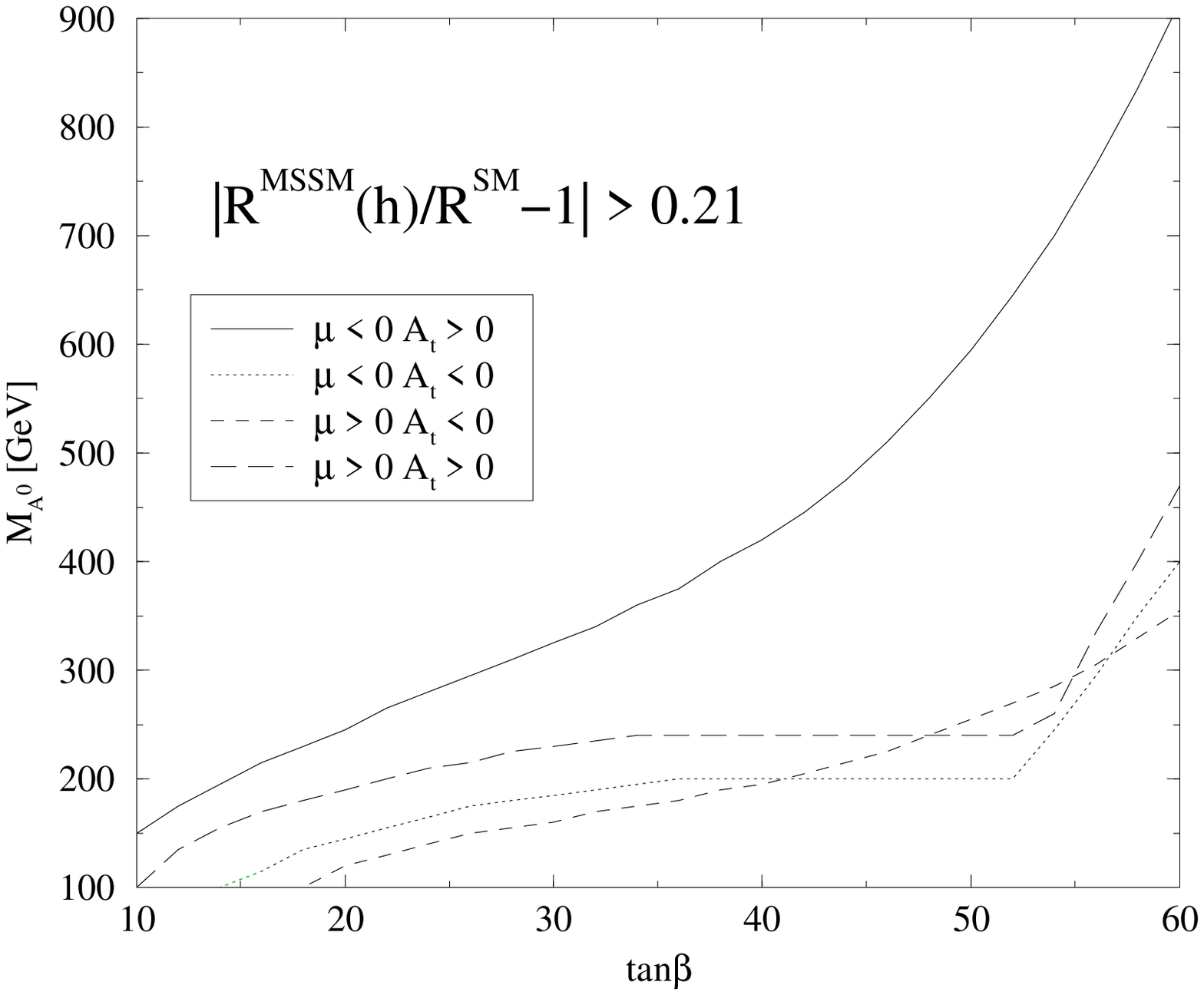}}\\
(a)&(b)
\end{tabular}
\end{center}\vspace*{-0.3cm}
\caption{Sensitivity regions on $\rmssm/\rsm$ with \textbf{a)} 5.4\%
  uncertainty in the measurement; \textbf{b)} 21\% uncertainty.}
\label{fig:exclregion}
\end{figure}

The decoupling behaviour with \ma\  becomes apparent in Fig.~\ref{fig:rmssmh}a.
We also clearly see in Fig.~\ref{fig:rmssmh} 
that $\rmssm(h)$ deviates significantly from the reference value
$\rsm$. In some favorable cases, i.e. small \ma, large \tb, $\mu<0$
and $A_t>0$,  the 
ratio (\ref{eq:Rh0}) can be as large as two. 
Clearly, a moderate-precision
measurement of this quantity would give clear signs of a Higgs boson
belonging to a SUSY model. For the LHC we estimate that this quantity
can be measured to a $21\%$ accuracy. 
By looking at the associate $WW$-fusion Higgs boson production
$qq\rightarrow W^*W^*\to H$, the 
$BR(H\to \tau^+ \tau^-)/BR(H\to \gamma \gamma)$ is measurable with an accuracy
of order $15\%$~\cite{Zeppenfeld}. On the other hand, for the associated
Higgs-boson production with a top quark
$(pp\to t\bar t H)$ the ratio $BR(H\to b\bar{b})/BR(H\to\gamma\gamma)$
can be performed with a similar precision~\cite{Gianotti}. From these
two independent 
measurements one determines $R$ with the error quoted above. If one were
able to make both measurements using the same Higgs-boson production
process, the error might be decreased.
The $\pm21\%$ deviation region is marked as a light-shaded region in the
figures. As for a future $e^+e^-$ Linear Collider (LC) running at $500\GeV$
center-of-mass energy, the simulation shows that the ratio of the 
effective Yukawa  couplings, $h_b/h_\tau(\equiv\sqrt{R})$, 
can be measured with an
accuracy of $2.7\%$~\cite{TESLATDR}. The corresponding band of $\pm
5.4\%$ accuracy in~(\ref{eq:Rh0}) is shown as a dark-shaded region. 

\begin{figure}[t]
\centerline{\resizebox{6cm}{!}{\includegraphics{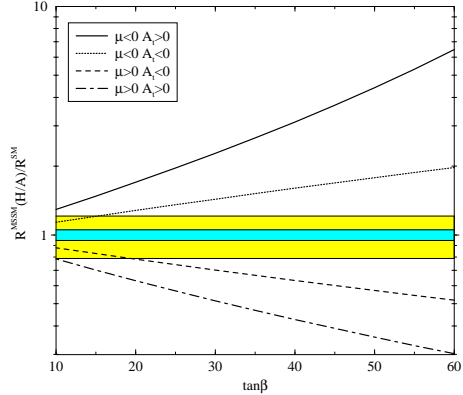}}}
\caption{Deviation of $\rmssm(H/A)$ with respect to the SM value, as a
  function of \tb\  for various choices
  of the SUSY parameters. The shaded regions are as in
  Fig.~\ref{fig:rmssmh}.}
\label{fig:rmssmH}
\end{figure}

We can now find the regions in the $(\tb,\ma)$ plane in which each
experiment can be sensitive to the SUSY nature of the lightest Higgs
boson. We show these regions in Fig.~\ref{fig:exclregion}a for a $5.4\%$
accuracy measurement, and in Fig.~\ref{fig:exclregion}b for a $21\%$ one.
We see that with a 5.4\% measurement one can have sensitivity to SUSY
for \ma\  up to $\sim 1.8\TeV$ in the most favorable scenario. In
less-favored scenarios the sensitivity is kept up to   
$\ma\sim800\GeV$, but  there exists also large regions where one is
sensitive to SUSY only up to $\ma\sim500\GeV$. 
However, all these masses are
well above the threshold production of the heavy Higgs particles for a
$500\GeV$ LC. We stress once 
again that these conclusions are independent of the scale of the SUSY
masses. As long as a $21\%$ accuracy is concerned, feasible e.g.\ at
the LHC, the regions sensitivity are of course much smaller
(Fig.~\ref{fig:exclregion}b). In this case one can probe the SUSY
nature of the Higgs boson only if $A^0$ is lighter that
$\sim900\GeV$. This means that the heavier MSSM Higgs bosons $H^0$, $A^0$ and
$H^\pm$ will also be 
produced at high rates at the LHC. Then, it would be more useful to
move our attention to $\rmssm(H/A)$ (corresponding
eqs.~(\ref{eq:DeffCoupH}), (\ref{eq:AnaliR})). We have checked that this 
quantity is very insensitive to $\tan\alpha$, and so to \ma. Its
numerical value is very close for both types of heavy neutral Higgs
bosons. We show the result of this analysis in
Fig.~\ref{fig:rmssmH}. A deviation of 21\% with respect to the SM value
is guaranteed for any scenario with $\tb\gsim20$;
hence, the SUSY
nature of the Higgs sector can be determined with a  
moderate-precision measurement.

To summarize, we have proposed the observable $R=\brHbb/\brHtt$ to
discriminate between SUSY and non-SUSY Higgs models. This observable
suffers  only little from systematic uncertainties, 
and is a theoretically
\textit{clean} observable.  In the MSSM, $R$ is affected by 
 quantum contributions that do not decouple 
even in the heavy SUSY limit. By assuming a $\pm5.4\%$
measurement of this ratio for the lightest Higgs boson, to be made at a
$500\GeV$ LC, one is sensitive to the SUSY nature of the lightest Higgs
boson $h^0$ for values of the $A^0$
mass up to $1.8\TeV$.  A less precise measurement at
$\pm21\%$ accuracy, feasible at the LHC,
is sensitive to SUSY only if
$\ma<900\GeV$. In this latter case the measurement of $R$
for the heavy 
Higgs bosons $A^0$ and $H^0$ is possible and  can give clear evidence
for, or against, the SUSY nature of the Higgs bosons. 
Further confirmation can be obtained by correlating these 
measurements with the 
production cross-section of charged Higgs bosons~\cite{Belaetal}.
Further simulation
analysis of the expected experimental determination are highly desirable.

\bigskip
 
\noindent 
We are thankful to D. Zeppenfeld for valuable comments.

\end{document}